\documentstyle[aps,preprint,tighten,epsfig]{revtex}
\draft

\def \bfgr #1{ \mbox {{\boldmath $#1$}}}
\newcommand{\be}{\begin{eqnarray}}
\newcommand {\CM} {{\cal M}}
\newcommand {\CF} {{\cal F}}
\newcommand{\ee}{\end{eqnarray}}

\def\bp{{\mbox{\boldmath$p$}}}

\def\bxi{{\mbox{\boldmath$\xi$}}}

\begin{document}

\title{Tagging the \boldmath {$p n \to d \phi$}
reaction by backward protons in
\boldmath{ $p d\to  d \phi p_{sp}$} processes}

\author{
{\sc L.P. Kaptari}$^{a,b,c}$,
{\sc B. K\"ampfer}$^a$, 
{\sc S.S. Semikh}$^{a,b}$ }

\address{
$^a$ Forschungszentrum Rossendorf, PF 510119, 01314 Dresden,
Germany\\
$^b$ Bogolyubov Laboratory of Theoretical Physics, JINR,
Dubna 141980, Russia\\
$^c$ Department of Physics, University of Perugia, and
INFN Sezione di Perugia, via A. Pascoli, I-06100, Italy }

\maketitle

\begin{abstract}
The reaction  $p d \to d  \phi p_{sp}$ is studied 
within the Bethe-Salpeter formalism.
Under special kinematical conditions (slow backward spectator proton
$p_{sp}$ and fast forward deuteron) relevant for forthcoming 
experiments at COSY, the  cross section and a set of polarization observables 
factorize in the contribution of the pure subprocess $pn\to d\phi$
and a contribution stemming from deuteron quantities and kinematical factors.
This provides a theoretical basis for studying threshold-near processes 
at quasi-free neutrons.\\[3mm]
{\it PACS:} 13.75-n, 14.20.-c, 21.45+v\\
{\it Keywords:} hadron-induced intermediate-energy reaction,
meson production, few-body systems
\end{abstract}

\section{Introduction} 

Data of elementary nucleon reactions involving a neutron in the entrance channel
are scarce. One can use, for instance, a tagged neutron ($n$) beam \cite{IUCF}
impinging on a proton ($p$) target or a proton beam impinging on a nuclear target. 
In the latter case one needs models to deconvolute the $pn$ reaction. Here we are 
interested in threshold-near meson production. Over the last decade the 
data basis for the $pp$ reactions has been impressively improved \cite{report}.
Such a detailed knowledge of elementary hadron reactions is necessary for 
analyzing heavy-ion collisions. Threshold-near meson production
is also interesting for testing predictions of QCD in the 
non-perturbative region, provided e.g., by chiral perturbation theory.
While it seems to be straightforward to translate
cross sections for reactions of the type $p p \to p p M$ ($M$ is a meson) 
into the cross section $n n \to n n M$, the channel $p n \to p n M$ is less simple
accessible. Model calculations point to a non-trivial energy 
dependence of the ratio 
$\sigma (p p \to p p M)/ \sigma (n p \to n p M)$, e.g. for $\phi$ meson production
\cite{titov1}. Experiments showed, furthermore, a surprisingly large ratio 
$\sigma(p d \to p d \eta)/\sigma(p p \to p p \eta)$
\cite{IUCF,chiavassa}. 
Therefore, it
is desirable to test directly predictions for the $pn$ reactions.  

The spectator technique \cite{johansson,anke}
offers one possibility to isolate to a large extent the 
quasi-free reaction at a neutron when a proton beam impinges on a deuteron
target. Since the final state interactions  
in the $pn$ system differs for singlet and triplet states and can be large
at threshold, it is of great advantage to consider
the particular reaction $p n \to d M$, as the deuteron $d$ is in one well
defined state. Guided by this, in \cite{guide,Nakayama} the reactions
$p n \to d M $ with $M = \eta, \eta', \omega, \phi, a _0^+$ have been considered.

The aim of the present note is to analyze the complete reaction
$pd\to dMp_{sp}$ and to test to what extent the spectator
technique \cite{johansson,anke} really selects a quasi-free reaction
at the neutron in the deuteron target. In other words, we are going to
derive a factorization theorem showing kinematical conditions 
for which the total cross section $\sigma(p d \to d M p_{sp})$ 
factorizes in a part depending on the target deuteron
characteristics and a part $\sigma(p n \to d M)$. In doing so we select as a
sufficiently transparent example the $\phi$ production. The reasons are obvious:
(i) the elementary $\phi$ production amplitude is simple \cite{titov1},
(ii) the $\phi$ meson is interesting with respect to OZI rule \cite{Cassing}
and hidden strangeness in the nucleon \cite{titov1,Nakayama,titov_oh,titov2},
and (iii) the $\phi$ meson is important  with respect to both the inclusive 
$K^-$ production \cite{kampfer} and the $e^+e^-$ decay channel to be studied 
at HADES in near future \cite{hades}. An extension to $\rho$ and $\omega$
production goes along the same lines, but due to the role of $s$, $u$ 
channel baryon resonances \cite{titov_oh,titov3,nakayama2002}
the elementary amplitude of the subprocess $p n \to p n M$ is by far 
more complicate and needs separate consideration, as other mesons too.

Our paper is organized as follows. In section II we present the kinematics,
cross section and amplitudes of the complete reaction  $p d \to d \phi p_{sp}$.
Numerical results are exhibited in section III, where polarization observables
are discussed and compared with the ones of the quasi-free sub-reaction 
$p n \to d \phi$. We also derive selection rules. The factorization of the
cross section is considered in section IV. The summary can be found in
section V.

\section{kinematics, cross section, amplitudes} 

\begin{figure}[ht]
\centering
\epsfig{figure=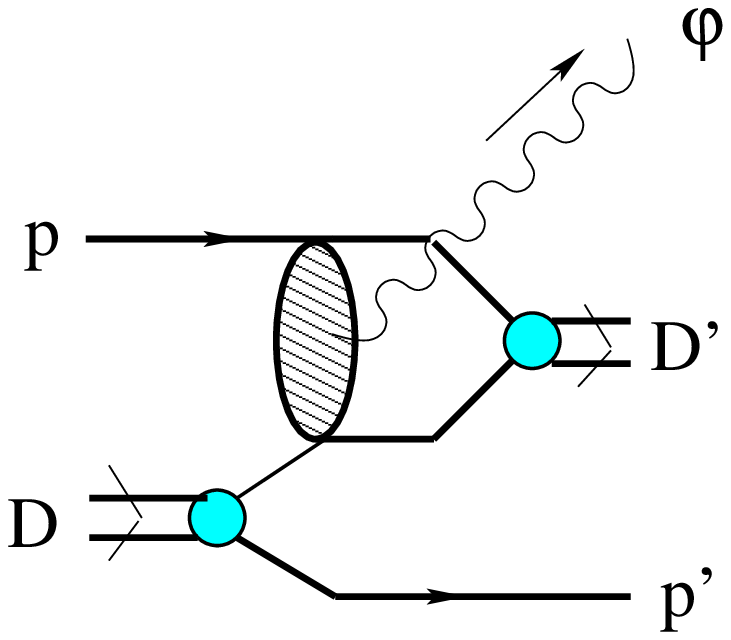,width=9cm}
\bigskip
\caption{Diagram of the process (\protect\ref{reaction}).}
\label{diagr1}
\end{figure}

Let us consider the process of $\phi$ meson production
in the exclusive reaction envisaged in \cite{anke}
\be
p_p + p_d = p_d' + p_\phi + p_p',
\label{reaction}
\ee
at kinetic energies of the incoming proton corresponding to those achievable at 
the Cooler Synchotron COSY $T_p \le  2.7$ GeV.
The deuteron is here supposed to be detected in the forward direction,
i.e., in direction of the incoming proton with relativistic energy,
$T_d' \sim 1$ GeV, and, contrarily, the final (spectator) proton is 
slowly moving in the backward direction 
($\theta_{sp} > 90^0$, $p_p' < 100$ MeV). 
The main contribution to the process (\ref{reaction}) comes then
from the spectator mechanism, where the incoming proton interacts with the
internal neutron from the deuteron and produces a meson and a deuteron in 
the final state. The second proton in the deuteron acts merely as a spectator. 
Schematically the spectator mechanism is depicted in Fig.~\ref{diagr1}.

In the laboratory system, where the target deuteron is at rest, 
the relevant momenta are defined by
$p_p=(E_p,\bp_p)$ and  $p_d=(M_d, {\bf 0})$ as initial momenta of
the proton and deuteron,
$p_p'=(E_p',\bp_p')$, $p_d'=(E_d',\bp_d')$, and $p_\phi=(E_\phi,\bp_\phi)$
as momenta of  final  proton, deuteron and
$\phi$ meson, respectively.
The fivefold differential cross section reads 
\be
2 E_{d}'\,\frac{d^5\sigma}{d^3p_{d}'\,d\Omega_{p}'} =
\frac{1}{8\,(2\pi)^5\sqrt{\lambda}}\,
\left [\frac{|\bp_p'|^2}{|B|\bp_p'|-C E_{p}'|}\right]\frac16
\sum\limits_{\it spins}
\,|\CF |^2,
\label{eq1}
\ee
where the K\"allen function $\lambda=4\, M_d^2\, |\bp_p|^2$ and
the kinematical coefficients $B=M_d+E_p-E_{d}'$ and
$C=|\bp_p|\, \mbox{cos}\,\theta_{p}' - |\bp_{d}'|\, \mbox{cos}\,\theta_{dp}'$
reflect the energy conservation.
The invariant amplitude $\CF$ depends, besides the spin projections
$s_p,\CM_d;s_{p}',\CM_{d}',\CM_\phi$
on the quantization z-axis (oriented along ${\bf p}_p$),
also upon the initial energy and five independent kinematical variables,
say  ${\bf p}'_d$ and $\Omega_{p}'$.

To compute the amplitude $\CF$, let us define the  scattering operator
${\hat{\cal O}}_{\CM_\phi,\CM_{d}'}$ (a $4\otimes 4$ matrix in the spinor space)
as the  truncated amplitude for the subprocess
$p n \to d \phi$ (hatched blob of the diagram in Fig.~\ref{diagr1}),
i.e., as an operator which, when sandwiched between
two free spinors $u_p$ and $u_n$, results in the amplitude of
the real process  $p n \to d \phi$ with two on-mass shell nucleons.
In the general case this operator is assumed to describe the process of 
meson production from two off-mass shell nucleons. With this definition,
the amplitude $\CF$ may be written as
\be
\CF
={\bar u}_\alpha(p_p',s_p')\,({\hat p}_p'-m)_{\alpha\beta}\,
\Phi_{\CM_d}^{c\beta}(p_n,p_p')\,
{\hat{\cal O}}_{\CM_\phi,\CM_{d}'}^{\delta c}\,u_\delta(p_p,s_p),
\label{eq3}
\ee
where $m$ is the nucleon mass and
$\Phi_{\CM_d}^{c\beta}(p_n,p_p')$ denotes the Bethe-Salpeter (BS)
amplitude of the deuteron in the momentum space. In Eq.~(\ref{eq3}),
the spinor indices are exhibited explicitly, with Latin and Greek characters
referring to neutrons and protons, respectively
(summation over indices occurring pairwise is supposed).
An expression for the amplitude (\ref{eq3}) can be found by specifying
a model for the operator  $ \hat{\cal O}$. However, in order
to keep the correspondence between the reaction
(\ref{reaction}) and the process of $\phi$ meson production in $NN$ 
collisions as close as possible, we
highlight explicitly its dependence upon the BS amplitude of the final deuteron
by introducing another operator $\hat O$ as
\be
{\hat{\cal O}}_{\CM_\phi,\CM_{d}'}^{\delta c}=
-i \int \, \frac{d^4p}{(2\pi)^4} \,
\bar \Phi_{\CM_d'}^{\alpha b}(1',2') \,
\hat O^{bc}_{\alpha\delta}(12;1'2',\CM_\phi).
\label{eq4}
\ee
Then from Eq.~(\ref{eq4}) and from Fig.~\ref{diagr1} it is easy to
recognize the operator  $\hat O(12;1'2',\CM_\phi)$
as the one describing the $\phi$ meson production in the sub-reaction 
$p n \to d \phi$ with an off-mass shell initial neutron.
Such an operator has been introduced
in Ref.~\cite{ourpnDphi} where its properties have been discussed in detail.
We briefly recall here the main steps in computing  $\hat O(12;1'2',\CM_\phi)$ and
the amplitude $\CF$ in Eq.~(\ref{eq3}). Usually, within the BS 
formalism in order to exhibit formulae
in a familiar matrix form, one introduces new, ``charge-conjugated''
quantities such as
$\Psi\equiv -\Phi \gamma_c$, ${\hat{\cal Q}}\equiv \gamma_c{\hat{\cal O}}$,
$\gamma_c\gamma_\mu^T=-\gamma_\mu\gamma_c$,
$v(p)=[{\bar u}(p)\gamma_c]^T$, ${\bar v}(p)=[\gamma_c u(p)]^T$.
Then the amplitude $\CF$ can be written as
\be
\CF
={\bar v}(p_p,s_p)\,{\hat{\cal Q}}_{\CM_\phi,\CM_{d}'}\,
\Psi_{\CM_d}(p_n,p_p')\,({\hat p}_p'+m)\, v(p_p',s_p'),
\label{eq5}
\ee
where in Eq.~(\ref{eq5}) and in subsequent expressions
one has usual matrix multiplications and traces. For further simplification
we insert  between ${\hat{\cal Q}}$ and $\Psi$ in
(\ref{eq5}) the complete set of the Dirac free spinors
$
{\hat I}=\left ( {1}/{2m}\right )\sum\limits_{r=1}^{4}\,\varepsilon_r\,
u_r(p){\bar u}_r(p)
$, where
$\varepsilon_r = +1$ for $r=1,2$ and $\varepsilon_r = -1$ for $r=3,4$,
to obtain  
\be
\CF
=\frac{1}{2m}\sum\limits_{s_n}T_{s_p,s_n}^{\CM_\phi,\CM_d'}(\nu_{pn},t)
\,\,{\bar u}(p_n,s_n)
\Psi_{\CM_d}(p_n,p_p')\,({\hat p}_p'+m)\, v(p_p',s_p').
\label{eq7}
\ee
The amplitude $
T_{s_p,s_n}^{\CM_\phi,\CM_d'}(\nu_{pn},t)
={\bar v}(p_p,s_p)\,
{\hat{\cal Q}}_{\CM_\phi,\CM_{d}'}\,u(p_n,s_n)
$
coincides with
the matrix element of the real process $p n \to d \phi$, defined by the
Mandelstam variables $\nu_{pn}=(p_p+p_n)^2$, $t=(p_p-p_d')^2$. Here is worth
emphasizing that the kinematics of the process (\ref{reaction}) differs
from the one of the real process $p n \to d \phi$. In spite of commonly defined 
variable $t$, the numerical values of momenta of the final meson and 
deuteron do not coincide in the two processes. Rather, the amplitude  
$T_{s_p,s_n}^{\CM_\phi,\CM_d'}(\nu_{pn},t)$
in the process (\ref{reaction}) corresponds to
an amplitude of meson production from an off-mass shell neutron. If the off-mass
shell effects were negligibly small, then 
$T_{s_p,s_n}^{\CM_\phi,\CM_d'}(\nu_{pn},t)$
in both processes would depend solely upon the momentum transfer $t$ and
the effective energy $\nu_{NN} = (E_p+E_p')^2-({\bf p}-{\bf p}')^2$.
As shown in Ref.~\cite{ourpnDphi}, $T_{s_p,s_n}^{\CM_\phi,\CM_d'}$ can
be written in the form
\be
T_{s_p,s_n}^{\CM_\phi,\CM_d'}(\nu_{pn},t)
=\frac{i}{(2m)^2}\sum\limits_{r,r'=1}^4
\int\,\frac{d^4p}{(2\pi)^4}\, A_{s_p,s_n; r, r'}^{\CM_\phi}(12;1'2')\,
\bar { v}_{r'}(2') \bar\Psi_{\CM_d'}(1',2')u_r(1'),
\label{eq81}
\ee
where the partial spin amplitudes $A_{s_p,s_n; r, r'}^{\CM_\phi}(12;1'2')$
can be calculated  within an effective one-boson exchange  model
by an explicit evaluation of  the  elementary diagrams depicted 
in Fig.~2 in \cite{ourpnDphi}.
We use the parameter set ``B'' in Ref.~\cite{titov2}
to calculate the amplitudes  $A_{s_p,s_n; r, r'}^{\CM_\phi}(12;1'2')$.
Now the second part of the matrix element in Eq.~(\ref{eq7}),
which includes the BS amplitude $\Psi_{\CM_d}$ for the deuteron at rest,
can be straightforwardly evaluated  by
using the explicit expressions for the $^3S_1^{++}$ and $^3D_1^{++}$ components
of the deuteron BS amplitude \cite{quad}. The result is
\be
\label{eq9}
\CF
= \sqrt{8\pi M_d}\,\sum\limits_{s_n}\,
T_{s_p,s_n}^{\CM_\phi,\CM_d'}(\nu_{pn},t)
\left[ u_S(|{\bf p}'|)\,C_{\frac{1}{2}s_p'\frac{1}{2}s_n}^{1\CM_d}
-u_D(|{\bf p}'|)\sqrt{4\pi}
\sum\limits_{m, \mu}C_{\frac{1}{2}s_p'\frac{1}{2}s_n}^{1\mu}
C_{2m1\mu}^{1\CM_d}\,Y_{2m}({\bf n}) \right],
\ee
where $M_d$ is the deuteron mass, $u_{S,D}$ denote
$S$ and $D$ waves the BS wave function \cite{quad,solution}, 
${\bf n}={\bf p}/|{\bf p}|$ and
$Y_{2m}({\bf n})$ the spherical harmonics of second order.
In deriving Eq.~(\ref{eq9}) we used the  identity 
$\sqrt{2}\sum\limits_{\nu_1\nu_2}C_{\frac12\nu_1
\frac12\nu_2}^{1 \CM_d}\,
\chi_{\frac12\nu_1} \chi_{\frac12\nu_2}^T=
i(\bfgr\sigma\bfgr \xi_{{\cal M}_d})\sigma_y$, where $C_{\frac12\nu_1
\frac12\nu_2}^{1 \CM_d}$ are the Clebsch-Gordan coefficients, 
$\chi_{\frac12\nu}$ and $\bxi$ stand for the spin-$\frac12$ functions
and the deuteron polarization vector, respectively.
Equation (\ref{eq9}) shows that an exact factorization of the cross section
into two parts (one depending upon the target
deuteron wave function, another one upon the kinematics and dynamics
of the subprocess $p n \to d \phi$) generally does not occur.
This is because the summation over the spins of the internal neutron $s_n$  
which leads to interference terms in the square of the matrix element. 
However, below we show that under special kinematical conditions an
approximate factorization is possible.

\section{Numerical results} 

From Eqs.~(\ref{eq7} - \ref{eq9}),
the differential cross section (\ref{eq1}) and any other polarization 
observables of the process $p d \to d \phi p_{sp}$ are calculable. 
The BS wave functions $u_{S,D}$ basically coincide with the 
non-relativistic $S$ and $D$ waves of the deuteron for small values
of the momentum of the spectator proton (see \cite{quad}).
Therefore, the relativistic and non-relativistic
expressions for the amplitude $\CF$ have the same form, and one may
perform also calculations by 
using deuteron wave functions obtained within a non-relativistic approach.
In our calculations we use the BS solution \cite{solution},
obtained with a realistic one-boson exchange  kernel. 
Additionally, Paris and Bonn wave functions are considered 
to investigate the sensitivity of observables on the deuteron model.

All our calculations have been performed within kinematical conditions corresponding
to the COSY Proposal \#75~\cite{anke}, i.e., the detected deuteron is supposed to be
in the forward direction and the spectator proton in
the backward hemisphere with $\theta_p' \sim 120^0 - 140^0$. The momenta of
the deuteron are chosen in the range $p_d' \sim 1.9$ - 3 GeV/c,
which means that the spectator proton is detected with low velocity, i.e.,
$p_p' \sim 0$ - 0.04 GeV/c. These kinematic conditions imply that both the
reaction (\ref{reaction}) and the subprocess $p n \to d \phi$ occur 
essentially near the threshold.

\begin{figure}[ht]
\centering
\epsfig{figure=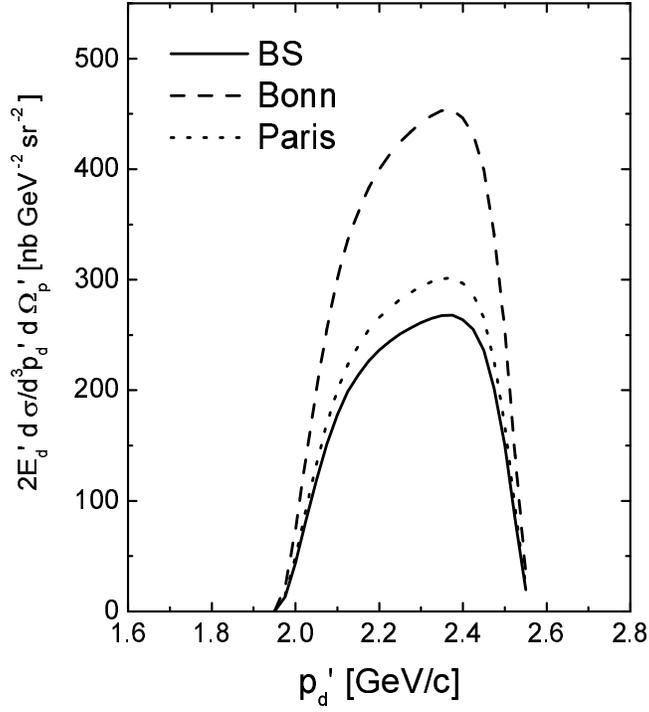,width=9cm}
\bigskip
\caption{Differential cross section of the reaction $p d \to d \phi p_{sp}$ as 
a function of the deuteron momentum $p_d'$ at $p_p = 3.505$ GeV/c, 
$\theta_{d'} = 0^o$,
$\theta_{p_{sp}'} = 130^o$.}
\label{pict3}
\end{figure}

In Fig.~\ref{pict3}, the differential cross section is presented as a function 
of the deuteron momentum for the specific choice of variables
$\theta_d' = 0^0$, and $\theta_p'=130^o$.
(We changed the kinematics with $\theta_d'$ up to $5^o$ and find
a very smooth behavior of all observable.) 
The kinetic energy of the initial proton is $T_p = 2.69$ GeV.
The maximum of the cross section at $p_d'\sim 2.35$ GeV/c
corresponds to that kinematical
condition where the momentum of the spectator proton is maximal, 
$p_{p}'\sim 0.035$ GeV/c.
Together with the large spectator angles this implies that
at these kinematical conditions the excess energy in the
$p n \to d \phi$ subprocess is minimal ($\Delta s^{1/2}\sim 4$ MeV)
and very near the threshold. Obviously, since at such kinematics the production amplitude
is almost constant, the maximum in the differential cross section
has a pure kinematical origin being entirely governed by
the kinematical factor within the square brackets
in Eq.~(\ref{eq1}). Fig.~\ref{pict3}
highlights that different models for the deuteron  wave function result in
quantitatively different cross sections. This is because  
the integration over the internal momentum in the final deuteron 
covers the region of the minimum of the $S$ wave where
different models provide essentially different wave functions (see 
discussion in \cite{ourpnDphi}).

Let us now discuss polarization observables.
Since the spin structure of the  amplitude $\CF$ in Eq.~(\ref{eq9})
is basically determined by  the spin structure of the amplitude
$T_{s_p,s_n}^{\CM_\phi,\CM_d'}(\nu_{pn},t)$ of the
subprocess $p n \to d \phi$, we proceed with an analysis of this amplitude.
Bearing in mind  that $T_{s_p,s_n}^{\CM_\phi,\CM_d'}(\nu_{pn},t)$ is
a Lorentz invariant quantity one can, for the sake of simplicity, 
investigate its structure
in the center of mass of the detected deuteron (near the threshold
this system coincides with the center of mass system of two nucleons),
provided the Lorentz boost is
performed along the $z$-axis. In such a case the spin structure of
$T_{s_p,s_n}^{\CM_\phi,\CM_d'}(\nu_{pn},t)$ is the same as in the laboratory 
system (no additional Wigner rotations are needed).
Near the threshold, the final $\phi-d$ system is in a $S$ orbital state
corresponding to $I_f = 0$, $J_D^\pi=1^+$
($L_D = 0,2$),  $J_f^\pi = J_\phi^\pi + J_D^\pi = 0^-,1^-,2^-$,
where $I$, $L$ and $J^\pi$ are
the total isospin, total radial angular momentum, angular momentum and parity,
respectively. Thus from symmetry constraints the allowed configurations  in the
initial state are $I_i = 0$,
$L_i = 1,3,5 \cdots$ and total spin $S_i = 0$.
The conservation law $J_i=J_f$ implies $L_i=1$, so that $J_i^\pi=1^-$.
Hence, near the threshold the main contribution into the amplitude
$T_{s_p,s_n}^{\CM_\phi,\CM_d'}(\nu_{pn},t)$ comes from  initial
spin configurations with $s_p = -s_n$ and, since the two nucleons
in the final state form a deuteron (with the total spin $S = 1$), the
most probable transitions near the threshold are expected to be those 
with nucleon spin-flips in the corresponding vertices and
with the deuteron projections $\CM_d' = \pm 1$. Consequently, the
projections of the $\phi$ meson are  to be $\CM_\phi=\mp 1$.

Coming back to the complete reaction one can expect that the prevailing
contribution into the full amplitude $\CF$ comes from configurations
with $\CM_d - s_p' = -s_p$, if 
(i) the effects of $D$ waves in the target deuteron may be disregarded 
near the threshold,
(ii) the spectator mechanism describes the reaction appropriately, and
(iii) the factorization (see below) holds. 
Hence, near the threshold the polarization observables of the reaction 
(\ref{reaction}) are predicted to behave in the same
manner as the ones of the subprocess $pn\to d\phi$. Therefore,
the study of various polarization observables allows to check 
the factorization.

Let us define the asymmetry of the reaction as
\be
&&
{\cal A} =\frac{d\sigma(s_p+s_n=1)+d\sigma(s_p+s_n=-1)-d\sigma(s_p+s_n=0) }
{d\sigma(s_p+s_n=1)+d\sigma(s_p+s_n=-1)+d\sigma(s_p+s_n=0)},
\label{assym}
\ee
where the spin of the intrinsic neutron is defined as $s_n=\CM_d-s_p'$. 
Then, from the above selection rules near the threshold, 
the asymmetry $A$ is predicted to be negative and close to -1. 
In Fig.~\ref{pict4}, results of a complete calculation of the asymmetry
$A$ are presented. The kinematical conditions are as in Fig.~\ref{pict3}. 
It is seen that in the whole kinematical range of $p_{d}'$ 
the asymmetry is basically the same as in the
subprocess $p n\to d \phi$. Moreover, as expected, at $p_d'$ near the maximum
the asymmetry approaches -1.

Similar results are obtained for other polarization observables.
For instance, the tensor analyzing power defined as
\be
&&
T_{20} =\frac{1}{\sqrt{2}}\frac{d\sigma(\CM_d'=1)+d\sigma(\CM_d'=-1)-2 d\sigma(\CM_d'=0) }
{d\sigma(\CM_d'=1)+d\sigma(\CM_d'=-1)+ d\sigma(\CM_d'=0) },
\label{t20}
\ee
is predicted to be $1/\sqrt{2}$ from the selection rules. Indeed, the full
evaluation of Eqs.~(\ref{eq7} -\ref{eq9}) yields a value close to that
(see Fig.~\ref{pict5}).

Apart from the polarization observables which near the threshold
are determined directly by the corresponding quantities for the
subprocess $pn\to d\phi$, one
can define another set of observables with completely different
behavior in the  two reactions, i.e., in the
subprocess $pn\to d\phi$ and in the reaction (\ref{reaction}).
For this purpose let us define observables at fixed
initial deuteron and proton  spin projections, e.g.,  at
$\CM_d = 0$ and $s_p = \frac12$, respectively. Near the threshold this means that
the spin projection of the neutron will be $s_n=-s_p'$ and by measuring the
spectator polarization one can form different combinations which determine
(allowed or forbidden) transitions for the subprocess $p n \to d \phi$.
Together with conservation rules for the complete reaction this admits
a study of different observables which vanish in the sub-reaction 
$p n \to d \phi$. Correspondingly, let us define the quantity
\be
&&
W_{zz} =\left.
\frac{d\sigma(\CM_\phi=1)-d\sigma(\CM_\phi=-1) }
{d\sigma(\CM_\phi=1)+d\sigma(\CM_\phi=-1) }\right |_{s_p=\frac12;\CM_d=0}.
\label{wzz}
\ee
If one measures this quantity for the positive spin  projection of the spectator,
then in the subprocess $p n \to d \phi$ one has super allowed transitions
($s_p+s_n =0,\, \CM_{\phi}=\pm 1,\, \CM_d'=\mp 1$) and, since in (\ref{wzz})
a summation over the final deuteron
projections is assumed, $W_{zz}$ is predicted to
approach zero at the threshold (see \cite{ourpnDphi}).

A different situation occurs if one considers $W_{zz}$ at $s_p'=-1/2$. In this case,
at least for $t=t_{min(max)} $ (i.e., the forward/backward
direction for the subprocess $p n \to d \phi$), the conservation
of the spin projection in the complete reaction (\ref{reaction})
strongly suppresses the
final states with $  \CM_{\phi}=\CM_{d}'=\pm 1 $ and
$\CM_{\phi} = - 1,\,\CM_{d}'=0$, so that the only contributing final states
remain these with $ \CM_{\phi}=1,\ \CM_{d}'=0,-1$ and $ \CM_{\phi}=-1,\ \CM_{d}'=1$,
which imply $W_{zz}\sim 1/3$, in contrast to the vanishing $W_{zz}$ 
in $p n \to d \phi$ \cite{ourpnDphi}.
Figure~\ref{pict6} illustrates our full numerical evaluation 
for $W_{zz}$ at $s_p'=-1/2$. 

\section{Factorization} 
   
\begin{figure}[ht]
\centering
\epsfig{figure=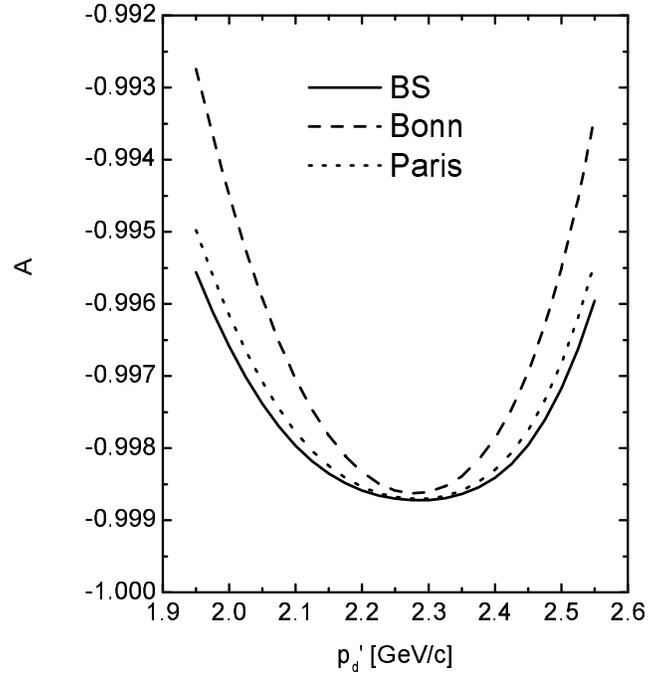,width=9cm}
\bigskip
\caption{As Fig.~\ref{pict3} but for the asymmetry $A$.}
\label{pict4}
\end{figure}

\begin{figure}[ht]
\centering
\epsfig{figure=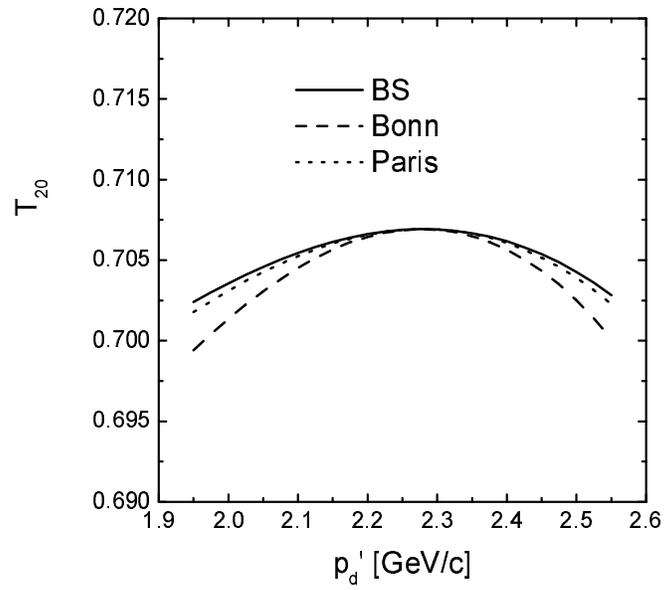,width=9cm}
\bigskip
\caption{As Fig.~\ref{pict3} but for $T_{20}$.}
\label{pict5}
\end{figure}

\begin{figure}[ht]
\centering
\epsfig{figure=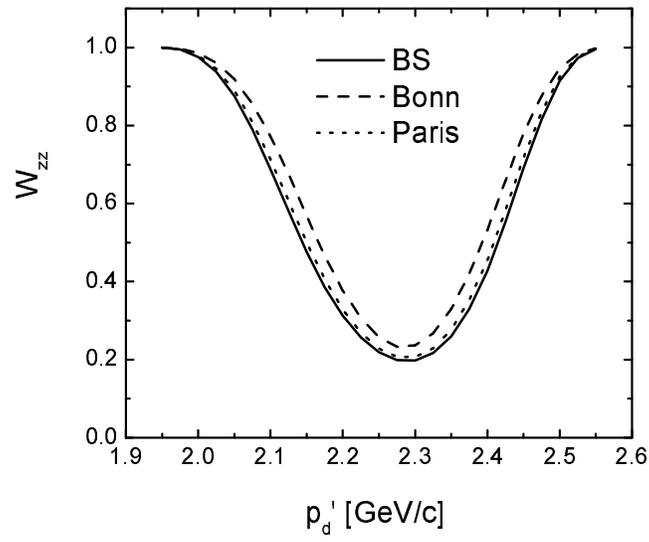,width=9cm}
\bigskip
\caption{As Fig.~\ref{pict3} but for $W_{zz}$.}
\label{pict6}
\end{figure}

Let us discuss in more detail the behavior of the cross section  
Eq.~(\ref{eq1}) with the matrix element (\ref{eq9}) near the threshold.
As mentioned above, the summation over the spin projection of the intrinsic
neutron $s_n$ in Eq.~(\ref{eq9}) prevents an exact factorization of
the cross section into the cross section of the subprocess $p n \to d \phi$ 
and a ``target factor''. However, 
from Figs.~\ref{pict4} - \ref{pict6} we conclude that, since the
deviations of the exactly calculated observables from those determined by the
selection rules of the subprocess $p n \to d \phi$ are small,
a factorization should hold indeed with a good accuracy.
For instance, the asymmetry in Fig.~\ref{pict4} demonstrates
the validity of the selection rule $s_p=-s_n$, i.e.,
the amplitude $T_{s_p,s_n}^{\CM_\phi,\CM_d'}(\nu_{pn},t)$ is
proportional to $\delta_{s_p,-s_n}$.
Similarly, Fig.~\ref{pict6} shows that the
main contribution comes from transitions with $|\CM_d'|=1$, 
and from the above selection rules the amplitude 
$T_{s_p,s_n}^{\CM_\phi,\CM_d'}(\nu_{pn},t)$ is
also proportional to $\delta_{\CM_d',-\CM_\phi}$. Hence, near the 
threshold one may represent the amplitude as
$
T_{s_p,s_n}^{\CM_\phi,\CM_d'}(\nu_{pn},t)  =
\delta_{\CM_d',-\CM_\phi}\,\delta_{s_p,-s_n} \,
\widetilde T_{s_p,\CM_d'}(\nu_{pn},t),
$
where $ \widetilde T_{s_p,\CM_d'}(\nu_{pn},t)$ is the
threshold-near amplitude of the process $pn\to d\phi$. Then in Eq.~(\ref{eq9})
the summation over $s_n$ disappears and, as conjectured above,
the unpolarized cross section
Eq.~(\ref{eq1}) factorizes. Substituting this amplitude 
into Eq.~(\ref{eq1}), introducing the corresponding flux factors for the
sub-reaction $p n \to d \phi$  and performing the summation over all spin projections,
the unpolarized cross section can be cast in the form
\be
2 E_{d}'\,\frac{d^5\sigma}{d^3p_{d}'\,d\Omega_{p}'} =
\frac{2}{(2\pi)^3}\,
\frac{\nu_{pn}(\nu_{pn}-4m^2)\,|\bp_p'|^2}{|\bp|\,|B|\bp_p'|-C E_{p}'|}\,
\left(u_S^2(|{\bf p}_p'|)+u_D^2(|{\bf p}_p'|)\right)\,
\frac{d\sigma (p n \to d \phi) }{dt},
\label{last}
\ee
where the differential cross section $d \sigma (p n \to d \phi) / dt$ 
depends only upon the transferred momentum
$t$ (determined by $p_p$ and $p_d'$) and the invariant mass $\nu_{pn}$ 
(determined by $p_p$ and $p_p'$ due to the balance $E_n = M_d - E_p'$,
$\bf p_n = - \bf p_p'$).
From Eq.~(\ref{last}) it is seen that by studying experimentally
the amplitude (\ref{eq9}), near the threshold,
one can obtain directly information about the subprocess
$pn\to d\phi$. Here it is worth reminding that the factorization of the
cross section holds only if the contribution of the $D$ wave is negligible, 
i.e., if in Eq.~(\ref{eq9}) one can put  $s_n=-s_p$
and arrives at
\be
\CF
\approx\sqrt{8\pi M_d}\,\,
T_{s_p,-s_p}^{\CM_\phi,\CM_d'}(\nu_{pn},t)
\left[u_S(|{\bf p}_p'|)\,C_{\frac{1}{2}s_p'\frac{1}{2}-s_p}^{1\CM_d}
-u_D(|{\bf p}_p'|)\sqrt{4\pi}
\sum\limits_{m\mu}C_{\frac{1}{2}s_p'\frac{1}{2}-s_p}^{1\mu}
C_{2m1\mu}^{1\CM_d}\,Y_{2m}({\bf n})\right].
\label{lastlast}
\ee\
Eqs.~(\ref{eq9}, \ref{last}) and (\ref{lastlast}) show that the smaller the
values of the spectator momentum are the higher the accuracy of the factorization
is. This has been confirmed numerically by comparing 
exact results based on Eqs.~(\ref{eq1}, \ref{eq9}) and
approximate results based on  Eqs.~(\ref{last}, \ref{lastlast}).
In the whole kinematical region of interests the difference is less than 0.5\%.

One may generally conjecture that a structure analog to Eq.~(\ref{eq9})
gives rise to factorization under suitable kinematical conditions,
hence providing a theoretical foundation of quasi-free reactions at the
neutron in the deuteron.

\section{Summary} 

In summary, within a covariant Bethe-Salpeter approach
we study the relation of the tagged quasi-free reaction $p n \to d \phi$
and the complete process $p d \to d \phi p_{sp}$ near the threshold.
Guided by the very proximity of the numerical results for cross section 
and polarization observables based on the exact formulae 
and on the factorized ones,
we find that, for special kinematical conditions, selection
rules are operative immediately causing a factorization 
$\sigma(p d \to d \phi p_{sp}) = F \ \sigma (p n \to d \phi)$ where $F$ depends
on deuteron and kinematical quantities. This is equivalent to an omission of the 
contribution of the deuteron $D$ wave at small values of the spectator momentum.

As an extension of the present work, an investigation of the 
twice tagged reaction
$d d \to n n M p_{sp 1} p_{sp 2}$ can be envisaged to study the quasi-free
meson production by the $n n$ reaction proposed in \cite{nn}. 

\acknowledgments 
We are grateful to A.I. Titov for many valuable discussions. L.P.K. and S.S.S.
acknowledge the warm hospitality of the nuclear theory group in the 
Research Center Rossendorf.
The work has been supported by Heisenberg - Landau program, 
BMBF grant 06DR921, and RFBR  00-15-96737.

\newpage 


\end{document}